\documentclass[aps,pra,twocolumn,showpacs]{revtex4-1}%
\usepackage{amsfonts}
\usepackage{amsmath}
\usepackage{amssymb}
\usepackage{graphicx}
\usepackage{dcolumn}
\usepackage{bm}
\usepackage{color}

\begin{document}
\title{Locking of length scales in two-band superconductors}

\author{M. Ichioka}
\affiliation{Department of Physics, RIIS, Okayama University, Okayama 700-8530, Japan}
\email{ichioka@okayama-u.ac.jp}
\author{V. G. Kogan}
\affiliation{Ames Laboratory, US Department of Energy, Ames, Iowa 50011, USA}
\author{J. Schmalian}
\affiliation{Institut f\"{u}r Theorie der Kondensierten Materie und Institut  f\"{u}r  Festk\"{o}rperphysik, Karlsruher Institut f\"{u}r Technologie, D-76131 Karlsruhe, Germany}


\begin{abstract}
A model of a clean two-band s-wave  superconductor with cylindrical Fermi surfaces, different Fermi velocities $v_{1,2}$,  and a general $2\times 2$ coupling matrix $V_{\alpha\beta}$ is used to study the order parameter distribution in vortex lattices. The   Eilenberger weak coupling formalism is used to calculate numerically  the spatial distributions of the pairing amplitudes $\Delta_1$ and $\Delta_2$ of the two bands for  vortices parallel to the Fermi cylinders. For generic values of the interband coupling $V_{12}$, it is shown that, independently of the couplings $V_{\alpha\beta}$, of the ratio $v_1/v_2$, of the temperature,  and  the applied field, the length scales of spatial variation of $\Delta_1$ and of $\Delta_2$ are the same within the accuracy of our calculations. The only exception from this  single length-scale behavior is found  for  $V_{12}\to 0$, i.e., for nearly decoupled bands. 
\end{abstract}

\date{\today}
\maketitle

\section{Introduction}
 
 Just at the dawn of the theory of multiband superconductors, it was established that near the critical temperature $T_c$, the coherence lengths, which set the length scales of spatial variation of the pairing amplitudes of the bands, are in fact the same, notwithstanding   differences in zero-$T$ BCS lengths $\xi_{0,\alpha}\propto v_\alpha/T_c$ ($\alpha$ is the band index and $v_\alpha$ is the Fermi velocity)~\cite{Geilikman}. This result has been ``rediscovered"  in the recent debate on the proper form of Ginzburg-Landau (GL) theory of two-band superconductors~\cite{Jani,KS}. This debate was triggered by extensive studies of multiband MgB$_2$ which prompted the formulation of two order-parameter  GL energy functionals to allow for different length scales $\xi_1\ne\xi_2$ associated with the two underlying bands, see \cite{Babaev,Babaev et al} and references therein. One of the predictions of these models was an intervortex attraction at distances large with respect to the London penetration depth. The observation of vortex clustering in MgB$_2$ in very small fields was considered as evidence for asymptotic intervortex attraction~\cite{Moschalkov}.  
 
 While it is established \cite{Geilikman,Jani,KS} that  near $T_c$, where the GL-expansion is justified, any generic superconductor with finite  interband coupling is governed by a single superconducting order parameter with one coherence length, it was pointed out in Ref. \cite{KS} that this does not have to be true away from $T_c$. Novel behavior is expected  especially in cases with different Fermi velocities of the  bands and for very weak interband coupling; this requires  to turn to microscopic descriptions of superconductors that are applicable at all temperatures. Interesting calculations of this kind  were performed in Refs. \cite{Silaev-Babaev,Milosevic} and showed that away from $T_c$  and for a very weak inter-band coupling the length scales $\xi_1$ and $\xi_2$ are indeed not equal, in particular  for low temperatures and at small magnetic fields. 
 
 However, there are several reasons why in real materials  the inter-band coupling is not as weak as was assumed in Refs. \cite{Silaev-Babaev,Milosevic}. First, the ever present Coulomb repulsion will inevitably give rise to off-diagonal matrix elements in band-representation, eventhough the usual renormalization of the Coulomb pseudopotential tends to reduce interband interactions more strongly than intraband interactions~\cite{MazinAntropov}. For MgB$_2$  the latter effect  was analyzed and is rather moderate \cite{MazinAntropov} : the bare interband Coulomb interaction is about half of the bare intraband interaction; renormalizations only reduce this ratio by another factor of 2, yielding interband Coulomb interactions that are approximately $25\%$ of the intraband couplings. Second, the matrix elements of the electron-lattice coupling within and between  electronic bands are for the important optical phonon branches a priori of the same order of magnitude.  Even for MgB$_2$, where the observation of a Leggett-mode in the Raman spectrum \cite{Blumberg07}  is evidence for comparatively weak interband coupling, a careful analysis of the inter- and intraband interactions reveals that the former is still about $20\%$  of the larger  and  similar to the smaller of the  intraband interactions~\cite{Liu01,Golubov02,Choi02a,Choi02b,MazinAntropov}. In other systems, such as the recently discussed iron-based superconductors is  even argued  that the interband coupling is the dominant source of pairing, see e.g. Refs. \cite{Mazin,KP}.  
 
 Further support for comparatively large interband coupling comes from an analysis of recent Scanning Tunneling Microscopy (STM) measurements of the density of states (DOS)  distribution within the vortex lattice at low temperatures in several two-band compounds~\cite{Suderow1,1144}.  For a single-band material one can construct  a phenomenological model to relate the measured zero-bias DOS distribution  $N(\bm r)$ to the pairing amplitudes $|\Delta(\bm r)|$ in the lattice unit cell~\cite{Suderow1}. This procedure is readily extended to a two-band situation, for which  $N(\bm r)$ depends on both $\xi_1$ and $\xi_2$.   The fit to the STM data for NbSe$_2$ and for NbSe$_{1.8}$S$_{0.2}$ showed that $\xi_1\approx \xi_2$  at $T=0.15\,$K$\,\ll T_c$. The same procedure has been applied to the novel  superconductor CaKFe$_4$As$_4$ with $T_c\approx 35\,$K and the zero-field tunneling spectrum having clearly two-gap features, again with the result $\xi_1\approx\xi_2$ at sub-Kelvin temperatures and at all fields examined~\cite{1144}.

 These theoretical considerations and observations   motivated us to re-examine  the question of the relative values of $\xi_1$ and $\xi_2$ in two-band superconductors within a microscopic approach that covers a broad temperature and magnetic field regime. In particular,  the analysis of the STM-data suggests that the emergence of one common length-scale is a much more robust phenomenon than one would expect for  moderately coupled multi-band system. Thus, we aim at clarifying the issue of when the coupling between two superconducting bands  becomes sufficiently strong to give rise  to a common length scale and under what conditions two separate length scales of the band-order parameters emerge. 

To this end, we use a ``brute-force" numerical procedure of solving Eilenberger equations for a vortex lattice in the two-band case  developed   in studies of MgB$_2$~\cite{Ichioka1}.    
 We consider a  weak-coupling model of a two-band superconductor with two Fermi surface parts having  different Fermi velocities and study the spatial variation of the pairing amplitudes $\Delta_{1,2}(\bm r)$ of the two bands within the vortex lattice unit cell. 
 While we analyze this model over a wide range of parameters, we do not focus on a specific application for a particular material. Rather, we intend to clarify general properties of the spatial dependency of $\Delta_{1,2}(\bm r)$.  Substantially different values of the Fermi velocities notwithstanding, the coherence lengths proportional to the vortex core size defined as $\xi^{(c)}_{1,2}\propto (d |\Delta_{1,2}|/d r)^{-1}_{r\to 0}$ ($r$ is the distance from the vortex center) turn out nearly the same for all choices of coupling constants $V_{\alpha\beta}$ examined ($\alpha,\beta=1,2$) except the case of nearly decoupled condensates  $V_{12}/ V_{11}\leq 0.1$. 
 
In  the limit $V_{12}\ll V_{11}$ our results agree with previous calculations \cite{Silaev-Babaev,Milosevic}. However, as soon as $V_{12}/ V_{11}\geq 0.1$, we obtain $\xi_1=\xi_2$, insensitive to details of coupling $V_{\alpha\beta}$, temperature, and field. Given the exponential dependence of the superconducting gap on the coupling constants, comparatively weakly coupled systems with $V_{12}/ V_{11}\geq 0.2\cdots 0.5$ may easily display interesting multiband behavior, such as collective fluctuations of the relative phase of the bands~\cite{Blumberg07}. However, our results show that  the system is nevertheless governed  by a single order-parameter  characterized by a single length scale.

\section{Approach}

We consider two-band system with two cylindrical Fermi surfaces ($\alpha=1,2$) both oriented parallel to the same crystal axis (the $c$ -axis) and  with Fermi velocities ${\bm v}_ \alpha({\bm k}) =v_ \alpha ( \cos\phi, \sin\phi ) $. $\bm k$ is the Fermi momentum and $\phi$  the corresponding azimuth.  The magnetic field is applied along $\bm c$ as well, i.e. the field is parallel to the axis of the cylinder.
For simplicity, the bands normal densities of states  are assumed  the same: $N_{0,1}=N_{0,2}=N_0$ (the total DOS per spin $N(0)=2N_0$). This assumption will not affect any of our results qualitatively and can easily be dropped. It still allows for distinct values of the Fermi velocities of the bands.  We set $v_{2} = 3 v_1$ to assure substantially different coherence lengths in the  limit  of fully decoupled bands. The $2\times 2$ coupling  matrix $V_{\alpha\beta}$ is assumed symmetric: $ V_{12}=V_{21}$.  
 
 Our approach is based on the quasiclassical version of the   weak-coupling BCS theory for 
  anisotropic Fermi surfaces and   order
parameters \cite{E}.  This theory is formulated in terms of  Eilenberger functions $f,\,\,f^+$ and $g $ (Gor'kov's Green's functions averaged over the energy): 
\begin{eqnarray}
&&(2\omega+ {\bm v}_\alpha\cdot{\bm \Pi})f_ \alpha=2\Delta_ \alpha g_ \alpha \,,
\label{Eil1}\\
 &&g_ \alpha^2=1-f_ \alpha f_ \alpha^{+}\,,\qquad  \alpha=1,2. \label{Eil3}
\end{eqnarray}
Here  ${\bm \Pi} ={\bm\nabla} +2\pi i{\bm A}/\phi_0$ with vector potential $\bm A$ and flux quantum $\phi_0$.
$ \omega=\pi T(2n+1)$ are  fermionic Matsubara frequencies with  integer $n$; hereafter $\omega$ and $T$ are  measured in energy units, i.e. $\hbar=k_B=1$.  The equation for $f^{+}$  is obtained from Eq.\,(\ref{Eil1}) by taking the complex conjugate and replacing ${\bm v}\rightarrow -{\bm v}$. 
  
The  pairing amplitudes satisfy the self-consistency relations: 
 \begin{eqnarray} &&
\Delta_ \alpha({\bm r})= 4\pi T N_0\sum_{\beta,\,\omega}  V_{\alpha\beta}  
\langle f( \omega,{\bm k},{\bm r})
\rangle _{\beta}\,,
\label{eq:scd}
\end{eqnarray} 
where the sum over positive Matsubara frequencies is extended up to $\omega_D$, the analog of Debuy frequency for electro-phonon mechanism;  
$\langle f( \omega,{\bm k},{\bm r})
\rangle _{\beta}$ stands for the average over the Fermi cylinder of the band $\beta$. 
  The contribution of the $ \alpha$-band to the current density is
\begin{eqnarray} && 
{\bm J}_ \alpha({\bm r})=-4\pi |e|N_0T\,  {\rm Im} \sum_{\omega >0} \langle 
{\bm v}g ( \omega,{\bm k} ,{\bm r}) \rangle_  \alpha\,,   
\label{eq:J}
\end{eqnarray} 
and the total current density is
\begin{eqnarray} &&
{\bm J} ={\bm J}_1  +{\bm J}_2={\bm\nabla}\times ({\bm \nabla}\times{\bm A})\,c/4\pi\,.
\label{eq:sca}
\end{eqnarray} 

The vector potential is taken in the form ${\bm A}({\bm r}) = ( {\bm B} \times {\bm r})/2 + \tilde{\bm A}({\bm r})$, where the magnetic induction $ {\bm B}=(0,0,B)$ is the field averaged over the vortex lattice cell and $\tilde{\bm A}({\bm r})$ represents the variable part of the field  which is periodic   in the vortex lattice and has zero spatial average. 
The unit vectors of the triangular vortex lattice are chosen as ${\bm u}_1=(a_0,0,0)$ and ${\bm u}_2=(\frac{1}{2}a_0,\sqrt{3}a_0 /2,0)$, where  the intervortex spacing is $a_0=(2\phi_0/\sqrt{3}B)^{1/2}$. 
We use periodic boundary conditions for the unit cell of the vortex lattice and take into account the order parameter phase winding around each vortex \cite{IHM}. 
  
Throughout the paper, we use Eilenberger units for the first band if it would have been single   ($V_{12}=V_{22}=0$):   $R_1=\hbar v_1/2\pi  T_{c1}$ is taken as a unit length ($R_1\approx 0.88\,\xi_{01}$ where $\xi_{01}$ is the zero-$T$ BCS coherence length of the ``bare" first band). 
  Fermi velocities are normalized to $v_1$, the magnetic field is measured in units of $B_1=\phi_0/2\pi R_1^2$ and the current density in $cB_1/4\pi R_1$, the energy unit is $\pi  T_{c1}$, and 
 $T_{c1}$ is the transition temperature in the single-band limit.
In these units, Eqs.\,(\ref{Eil1}) and (\ref{eq:J}) take the  form:
\begin{eqnarray}
&&( \omega+ {\bm v}_ \alpha\cdot{\bm \nabla})f_ \alpha= \Delta_ \alpha g_ \alpha -  i {\bm v}_\alpha\cdot [( {\bm B} \times {\bm r})/2 + \tilde{\bm A} ]f_ \alpha\,,\qquad
\label{Eil1a}\\
 &&  
{\bm J}_ \alpha({\bm r})=-\frac{2T}{\kappa_1^2} \sum_{\omega>0} 
\langle {\bm v} \, {\rm Im}\, g(\omega,{\bm k},{\bm r}) 
\rangle_  \alpha\,. \quad 
\label{eq:Ja}
\end{eqnarray}
Hereafter we keep the same notation for dimensionless quantities as for their dimensional counterparts; we will indicate explicitly if common units are needed.  

 The quantity $\kappa_1=\phi_0T_{c1}/\pi \hbar^2v_1^2\sqrt{2N_0}$ has   the same order of magnitude as the GL parameter   for one-band isotropic case, 
$\kappa_{GL}=3\phi_0T_{c}/ \hbar^2v^2\sqrt{7\zeta(3)N(0)}$. However,  $\kappa_1$ does not have the meaning of the penetration-depth-to-coherence-length ratio for the two-band system\cite{Jani,KS}, rather it is a convenient dimensionless  material parameter.

The dimensionless self-consistency  equations take the form:
 \begin{eqnarray} &&
\Delta_ \alpha(\bm r)= 2t N_0V_{11}\sum_{\beta,\,\omega}  \frac{V_{\alpha\beta}}{V_{11}} 
\langle f( \omega,{\bm k},{\bm r})\rangle _{\beta}\,,\label{eq:sce}\\
&&\pi e^{-\gamma}T_{c1}=2\omega_D\exp(-1/N_0V_{11})\,,\quad t= T/T_{c1}   
\label{Tc1} 
\end{eqnarray} 
where $\gamma$ is the Euler constant. In our calculations we set the cutoff frequency $\omega_D=40\, T_{c1}$ and $\kappa_1=4$. The numerical procedure is outlined in the Appendix.

The   profiles of the pairing amplitudes $|\Delta_\alpha({\bm r})|$  in real space are fitted by a $5^{\rm th}$-order polynomial near the vortex center along the nearest neighbor vortex direction.  We estimate the vortex core size $\xi^{(c)}_{\alpha}$ from  
\begin{eqnarray} 
\Delta_ \alpha({\bm r}) = \Delta_{{\rm m}, \alpha} \frac{r}{\xi^{(c)}_ \alpha} +O(r^2) \,, 
\quad  j=1,2 
\end{eqnarray}
near the vortex center. 
$\Delta_{{\rm m}, \alpha}$ is the maximum value of $|\Delta_ \alpha({\bm r})|$ within the unit cell.

\section{  ${\bm V_{12}}$   of the same order as ${\bm V_{11}}$} 

First, we present our results for   $V_{12}=0.32 V_{11}$. 
In order to see the effect of the coupling  in the second band,  
we consider two cases: $V_{22}=0$ and $V_{22}=0.32V_{11}$. 

\begin{figure}
\begin{center}
\includegraphics[width=7.0cm]{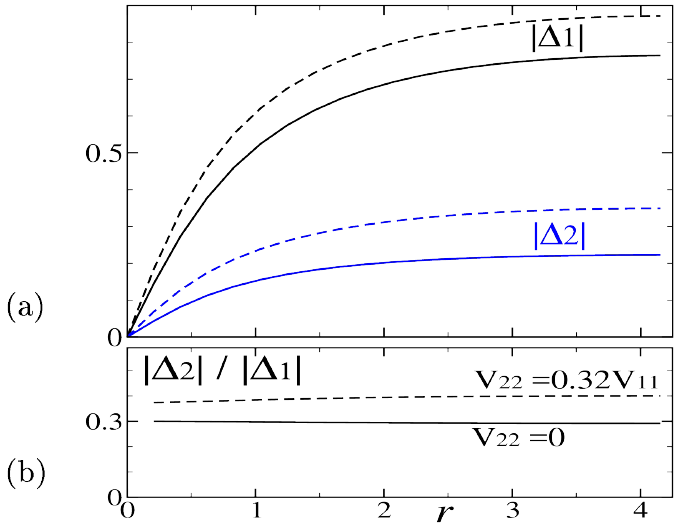}
\end{center}
\caption{(Color online)
(a) Pairing amplitudes $|\Delta_1({\bm r})|$ and $|\Delta_2({\bm r})|$ (in units $\pi T_{c1}$) vs distance $r$ (in units of $R_1=\hbar v_1/2\pi  T_{c1}$) from the vortex center to the midpoint between nearest neighbor vortices. 
In this calculation,  $V_{12}/V_{11}=0.32$, $t=T/T_{c1}=0.5$, and $B=0.1$ (in units $\phi_0/2\pi R_1^2$). 
Solid lines are for $V_{22}=0$,   dashed lines  are for $V_{22}/V_{11}=0.32$. (b) Nearly constant ratios $|\Delta_2({\bm r})|/|\Delta_1({\bm r})|$  imply  the same length scales for both pairing amplitudes.
}
\label{fig1}
\end{figure}

The profiles of $|\Delta_1({\bm r})|$ and $|\Delta_2({\bm r})|$ are shown in   Fig.\,\ref{fig1}(a). 
Near the vortex center, both $|\Delta_1({\bm r})|$ and $|\Delta_2({\bm r})|$ recover over the same lengths; this is seen  most directly in   panel (b) where nearly constant ratios $|\Delta_2({\bm r})|/|\Delta_1({\bm r})|$ are shown.  
In the presence of finite intraband coupling of the second band  $V_{22}$, the amplitude of  the pair potential of this band increases, with  $|\Delta_2({\bm r})|/|\Delta_1({\bm r})| \sim 0.4$, as expected. The spatial dependence of the two pair potentials is however the same.

\begin{figure} 
\begin{center}
\includegraphics[width=7.0cm]{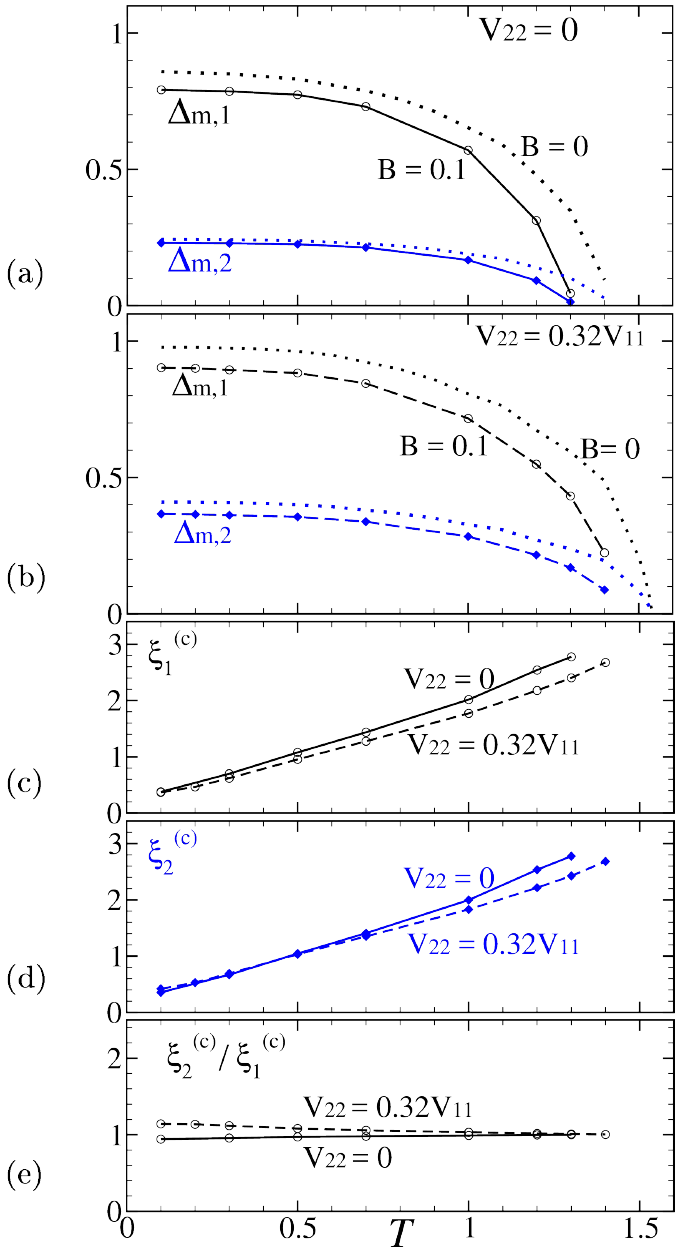}
\end{center}
\caption{(Color online)
(a) Temperature  dependence of maximum values $\Delta_{{\rm m},\alpha}$ of 
pairing amplitudes $|\Delta_\alpha({\bm r})|$ at $ B=0.1$ for $V_{22}=0$ and $V_{12}=0.32\,V_{11}$.  
Zero-field $|\Delta_\alpha|$ are  shown by dotted lines.  
(b) The same as (a) for  $V_{22}=0.32V_{11}$. 
(c,d) $T$ dependences of   core sizes $\xi^{(c)}_\alpha$, 
and (e) of $\xi^{(c)}_2/\xi^{(c)}_1$ for $B=0.1$.
Temperature $T$ is in units of $T_{c1}$.}
\label{fig2}
\end{figure}

Temperature dependences of the core radii $\xi^{(c)}_ \alpha$ and of the maximum value $\Delta_{{\rm m}, \alpha}$  are given in Fig.\,\ref{fig2}. 
While $\Delta_{{\rm m}, \alpha}$ are slightly smaller than those in zero field (dotted line) as they should,   the $T$-dependence of $\Delta_{{\rm m},\alpha}$ is similar to that at zero field. Nearly constant  ratios $\Delta_{{\rm m},2}/\Delta_{{\rm m},1}$ are  $\approx 0.3$ for $V_{22}=0$  and $\approx 0.4$ for $V_{22}=0.34V_{11}$. 
As the temperature increases, this ratio changes little: from  $0.291$ to  $0.295$ for $V_{22}=0$, and from  $0.406$ to $0.392$   for $V_{22}=0.32V_{11}$, respectively.   Within our analysis  we also reproduce Kramer-Pesch shrinking of the vortex core sizes $\xi^{(c)}$ on cooling  \cite{KramerPesch,Ichioka1996,Gumann}, see Fig. \ref{fig2}(c,d). 
 Thus,   we obtain  $\xi^{(c)}_2 \approx \xi^{(c)}_1$ in the whole temperature range. While it is expected \cite{Geilikman,Jani,KS} that $\xi^{(c)}_2/\xi^{(c)}_1 \rightarrow 1$ for $T\rightarrow T_c$, our finding of numerically very similar length scales over a broad temperature regime is rather surprising.

\begin{figure}
\begin{center}
\includegraphics[width=7.0cm]{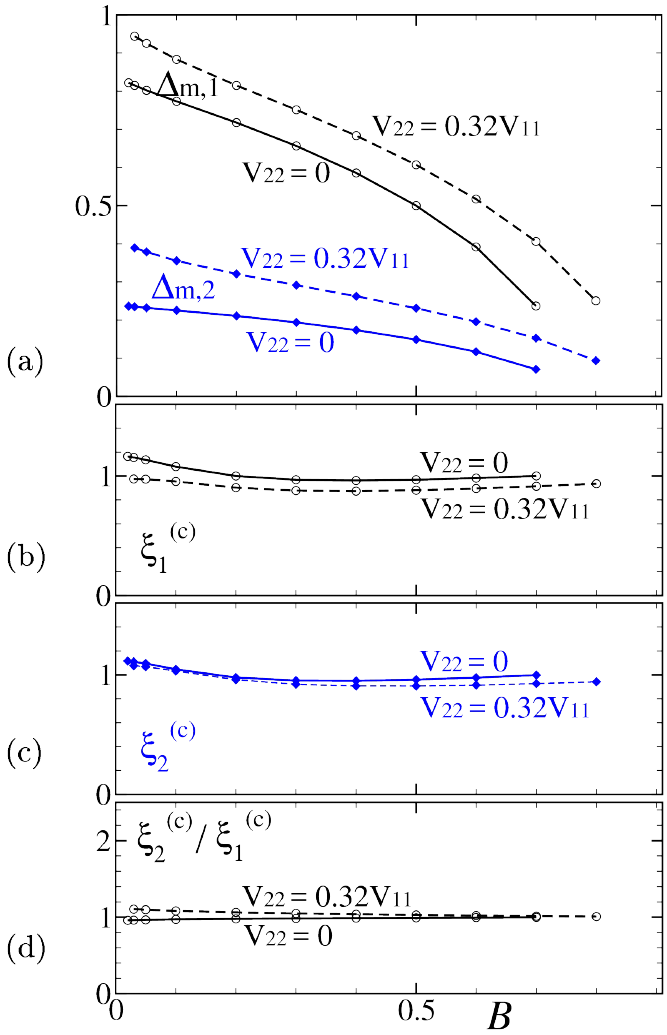}
\end{center}
\caption{(Color online)
(a) Magnetic field dependence of $\Delta_{{\rm m},\alpha}$, $\alpha=1,2$.  
(b,c) $ B $ dependence of the core sizes $\xi^{(c)}_\alpha$  and (d) the ratio $\xi^{(c)}_2/\xi^{(c)}_1$. Inputs:  
$t=0.5$,  $V_{12}=0.32V_{11}$,  
solid  lines are for $V_{22}=0$, dashed lines for  $V_{22}=0.32V_{11}$. 
}
\label{fig3}
\end{figure}

The field dependencies  of the pairing amplitudes and deduced length scales are shown in Fig.\,\ref{fig3}. 
 As expected, the $\Delta_{{\rm m}, \alpha}$ are suppressed upon increasing  the magnetic field,  see Fig.\,\ref{fig3}(a).
  As shown in  Fig.\,\ref{fig3}(b,c), after a slow decrease at low $ B$'s, the core radii $\xi^{(c)}_ \alpha$ are once again nearly constant over a wide range of field values. Most importantly however,  we find at all fields   that  $\xi^{(c)}_1 \approx \xi^{(c)}_2$,  see panel (d) of Fig.\,\ref{fig3}.
As $B$ approaches the upper critical field $H_{c2}$, $\xi^{(c)}_2/ \xi^{(c)}_1 \to 1$, see Fig.\,\ref{fig3}(d). This conclusion agrees with  the two-band theory of $H_{c2}$  \cite{KP-ROPP}, where it has been shown that near a 2$^{\rm nd}$ order phase transition at $H_{c2}$, the two pairing amplitudes satisfy the system of equations $-\xi^2{\bm \Pi}^2\Delta_ \alpha=\Delta_ \alpha$ with the same $\xi$. 

\section{Decoupling limit ${\bm V}_{12}\ll {\bm V}_{11}$  } 

Next we analyze the regime of almost decoupled band. In this limit, the two superconducting condensates are nearly independent.  
The vortex core radii can be different and dependent on the characteristics of the bands \cite{Silaev-Babaev,Milosevic}. 

\begin{figure} 
\begin{center}
\includegraphics[width=7.0cm]{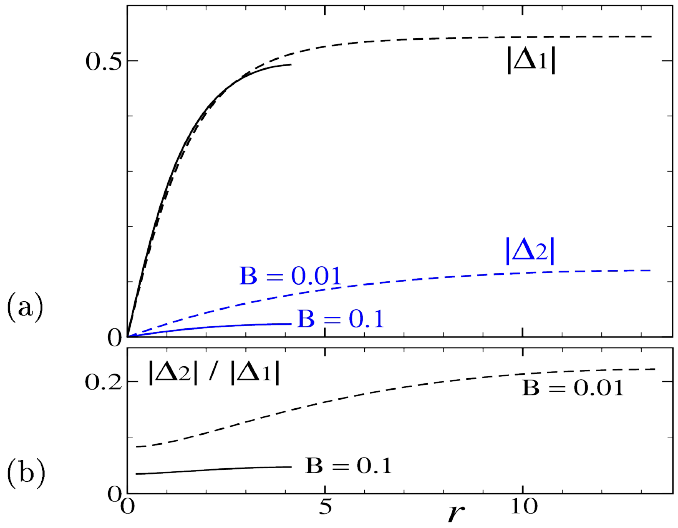}
\end{center}
\caption{(Color online)
(a) $|\Delta_1({\bm r})|$ and $|\Delta_2({\bm r})|$ vs   distance $r$ from the vortex center to the midpoint between nearest neighbor vortices. 
(b)  $|\Delta_2({\bm r})|/|\Delta_1({\bm r})|$. 
 Input parameters are $V_{12}=0.01V_{11}$, $V_{22}=0.85V_{11}$, and $t=0.5$; 
solid  lines are for $B=0.1$, dashed lines are for  $ B =0.01$.
}
\label{fig4}
\end{figure}

We consider  a  weak inter-band coupling, $V_{12}=0.01V_{11}$, whereas  $V_{22}=0.85V_{11}$. 
The resulting $|\Delta_ \alpha({\bm r})|$ are presented in   Fig.\,\ref{fig4}(a). 
At a low field $B=0.01$ (dashed lines), the recovery of $|\Delta_2({\bm r})|$ with increasing $r$ is indeed slow compared to $|\Delta_1({\bm r})|$, and  as a result we find that $\xi^{(c)}_2>\xi^{(c)}_1$. 
This behavior can also be seen in the $r$ dependence of the ratio $|\Delta_2({\bm r})|/ |\Delta_1({\bm r})|$, which is no longer constant, but decreases near the vortex core,    see Fig.\,\ref{fig4}(b). 
For higher field, $ B=0.1$ (see the solid lines in Fig. \ref{fig4}), $|\Delta_1({\bm r})|$ within the core region  does not change substantially compared to the low-field case, whereas $|\Delta_2({\bm r})|$ is suppressed strongly, as the intervortex distance is too short for the recovery of $|\Delta_2({\bm r})|$. 
In other words, since the ``effective $H_{c2}$" of the second band is small due to a   larger coherence length ($  v_2=3v_1$ and  $\Delta_2$ is small), superconductivity of the second band is easily suppressed by magnetic fields. 
Hence, at  high fields, the contribution  to superconductivity of the  second band is weak.   

\begin{figure} 
\begin{center}
\includegraphics[width=7.0cm]{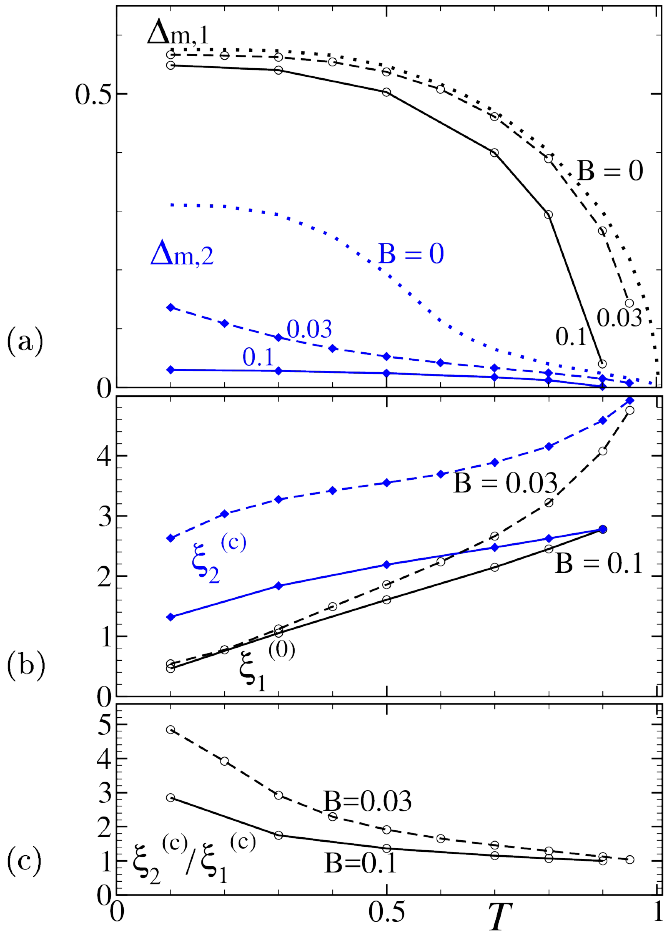}
\end{center}
\caption{(Color online)
(a) Temperature  dependence of $\Delta_{{\rm m},\alpha}$ at $B=0$ (dotted lines), $B=0.03$ (dashed lines), and $B=0.1$ (solid lines).  
(b) $T$ dependence of the vortex core radii $\xi^{(c)}_1$ and $\xi^{(c)}_2$, and (c) the ratio $\xi^{(c)}_2/\xi^{(c)}_1$. 
$V_{12}=0.01V_{11}$ and $V_{22}=0.85V_{11}$.
}
\label{fig5}
\end{figure}
\begin{figure} 
\begin{center}
\includegraphics[width=7.0cm]{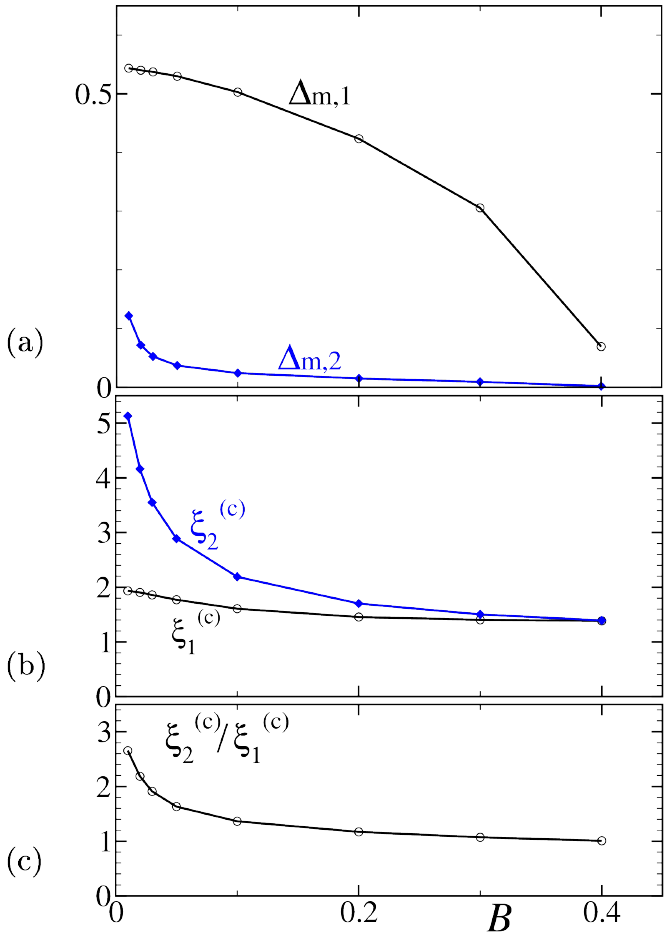}
\end{center}
\caption{(Color online) 
(a) The field   dependence of $\Delta_{{\rm m}, \alpha}$.  
(b) $B$-dependence of  core radii $\xi^{(c)}_1$ and $\xi^{(c)}_2$, and (c) the ratio $\xi^{(c)}_2/\xi^{(c)}_1$. Input parameters:
$t=0.5$, $V_{12}=0.01V_{11}$ and $V_{22}=0.85V_{11}$.
}
\label{fig6}
\end{figure}

The corresponding temperature dependence of the nearly decoupled band regime is shown in Fig.\,\ref{fig5}. 
$\Delta_{{\rm m},1}$ has the typical $T$-dependence of the BCS theory. 
However,  $\Delta_{{\rm m},2}(T)$ is different.  At low $T$, the superconductivity of the second band is enhanced, since it is caused here by $V_{22}=0.85\,V_{11}$.  For  $ B =0$, $\Delta_2$ is very small at elevated temperatures.  Above the intrinsic transition temperature of the decoupled second band,  superconductivity  of this band is only induced  by the weak interband coupling $V_{12}$, an observation that was made already shortly after the formulation of the BCS-theory \cite{Suhl59}.
With increasing $B$, the enhancement of $\Delta_{{\rm m},2}$ at low $T$ disappears and  practically vanishes at $B=0.1$. The $B$-dependence of the pairing amplitudes are shown in Fig. \ref{fig6}. $\Delta_{{\rm m},2}$   decreases rapidly at low $B$ reflecting small effective $H_{\rm c2,2}$ of the second band, and remains small at higher $B$ due to weak coupling $V_{12}$.   In the high $B$ range,   $\xi^{(c)}_1 \approx \xi^{(c)}_2$. This  combination of field and temperature variation of nearly decoupled bands may serve as a tool to identify whether one is indeed in this limit.
 
We note that the Kramer-Pesch 
shrinking of $\xi^{(c)}_2$ on cooling  is weak compared to that of $\xi^{(c)}_1$, see Fig.\,\ref{fig5}(b). 
Thus, the ratio $\xi^{(c)}_2/\xi^{(c)}_1$ increases upon lowering $T$. 
On the other hand, at higher $T$ and for fields approaching $H_{c2}$, $\xi^{(c)}_2/\xi^{(c)}_1 \to 1$ (again in agreement with Ref. \cite{KP-ROPP}).

\section{Discussion}

The issue of the  spatial variation of the superconducting order parameter in multi-band systems is interesting and relevant, in particular because of an increasing number of physical systems that clearly display multi-band behavior in their superconducting properties.  In addition to the description of  the variation of the order parameter near vortex cores, the  DOS distribution is related to $\Delta(\bm r)$ and is measurable. Recent STM low-$T$ data, interpreted within a phenomenological model, suggest that $\xi^{(c)}_1=\xi^{(c)}_2$ \cite{Suderow1}. While such length-scale locking is to be expected in the immeadiate vicinity of the transition temperature, it is not obvious away from $T_c$. Thus, a microscopic analysis of this open question is timely and  relevant. It shows that, within the accuracy of our numerical routines,  $\xi^{(c)}_1\approx \xi^{(c)}_2$ if the inter-band coupling   is of the same order as intra-band ones. This conclusion turns out to be valid at all temperatures and fields. In agreement with other microscopic calculations \cite{Silaev-Babaev,Milosevic}, we find this rule is violated  for a  very  weak inter-band coupling when the system is close to the limit of nearly decoupled  condensates.  The peculiar field and temperature dependence of such nearly decoupled bands can easily be used to test, for a given material, whether the coupling between bands is weak or only moderate.

\begin{figure} 
\begin{center}
\includegraphics[width=7.0cm]{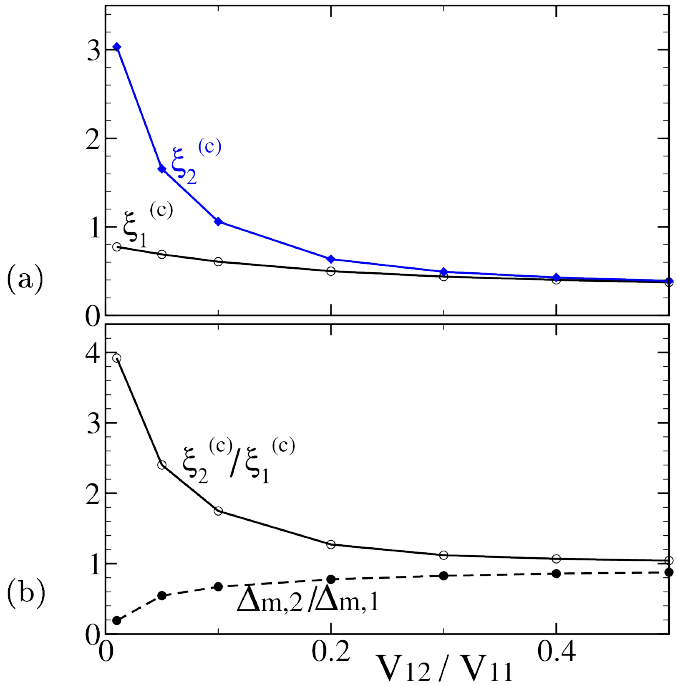}
\end{center}
\caption{(Color online) 
(a) $\xi^{(c)}_\alpha$ vs interband coupling $V_{12}/V_{11}$ for $V_{22}=0.83\,V_{11}$  at $t=0.2$ and $B=0.03$. 
(b) Ratios $\xi^{(c)}_2/\xi^{(c)}_1$ and $\Delta_{\rm m,2}/\Delta_{\rm m,1}$ vs $V_{12}/V_{11}$ for the same parameters as (a).
}
\label{fig7}
\end{figure}

To make this statement more quantitative, we show in Fig. \ref{fig7} the ratio  $\xi^{(c)}_2/\xi^{(c)}_1$  as a function of the inter-band coupling $V_{12}/V_{11}$ at  fixed   $t=T/T_{c1}=0.2$  and $B=0.03$. One sees that this ratio exceeds the value of 2 only when roughly $V_{12}/V_{11} < 0.1$. As discussed above, MgB$_2$ can be very well described by  $V_{12}/V_{11} \approx  0.2$ (see Refs. \cite{Liu01,Golubov02,Choi02a,Choi02b,MazinAntropov}). Thus, we conclude that this systems is not in the limit where two distinct characteristic length scales emerge.

In conclusion, by solving the quasi-classical Eilenberger equations, we analyzed the spatial variation of the pairing amplitudes within the vortex lattice of a two band superconductor over a wide range of temperatures and magnetic fields. Near the superconducting transition temperature $T_c(B)$,  
it is established~\cite{Geilikman,Jani,KS,KP-ROPP} that the emergence of one order parameter naturally implies that the spatial variation of this order parameter is governed by a single length scale. Away from $T_c$ it is however natural to expect that for a sufficiently weak coupling between the bands, distinct characteristic length scales  for the respective pairing amplitudes emerge. However, what precisely is meant by  {\em sufficiently weak}  has not been investigated thus far. Here we showed that such decoupling of the length scales occurs for values of the interband pairing interaction $V_{12}$ that is less than one order of magnitude of the largest intraband coupling. For larger values of the interband coupling a common temperature variation of the length $\xi^{(c)}_1$ and $\xi^{(c)}_2$  of the pairing amplitudes sets in. What is most surprising about our results is that these two length scales not only follow a common $T$-dependence, they are essentially identical in their magnitude $\xi^{(c)}_1 \approx \xi^{(c)}_2$. In other words, we observe a robust length scale locking of moderately coupled multiband superconductors.  Whatever difference might there be in the values of the length scales of the uncoupled system, our analysis shows that this difference is most likely to disappear everywhere below $ H_{c2}(T)$.  

In this work we only considered the situation of a clean two-band situation. Usually, the impurity scattering 
is expected to cause an isotropization of superconducting characteristics. 
Hence, we do not expect scattering to amplify differences of the length scales  $\xi_\alpha$. Still, as discussed in Ref.\,\cite{KP},    inter-band scattering can cause the superconductivity to become gapless with two bands acquiring substantially different DOSs in superconducting state. The question of how this difference   affects $\xi_\alpha\,$ remains to be answered.

\section*{Acknowledgements} 

We thank Lev Boulaevskii for illuminating comments. Work of V.K. was supported by the U.S. Department of Energy, Office of Science, Basic Energy Sciences, Materials
Sciences and Engineering Division. The Ames Laboratory is
operated for the U.S. DOE by Iowa State University under
Contract No. DE-AC02-07CH11358.  
 
\appendix

\section{Numerical method}

\begin{figure}[t] 
\begin{center}
\includegraphics[width=7.0cm]{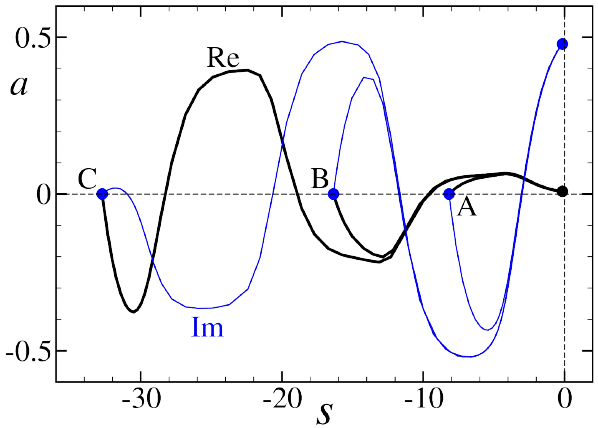}
\end{center}
\caption{(Color online) 
Solving  the first-order ordinary differential Eq.\,(\ref{eq:Ric1}) 
along the trajectory ${\bm r}'={\bm r}+s \hat{\bm v}_\alpha$ 
for  $a$ at $s=0$. 
Real and imaginary part of $a$ are shown for start positions $s_0=-8.2$ (A), $-16.4$ (B) and $-32.7$ (C). 
It is seen, that $a$  converges to the same solution at $s=0$. 
Input parameters are $\phi=1.25^\circ$ for ${\bm k}$, $\alpha=1$, 
$\omega=\pi T$ and $V_{22}=0$ in the case of Fig.\,1. 
${\bm r}$ is near the midpoint $(-a_0/2,0)$ between nearest neighbor vortices. 
}
\label{fig8}
\end{figure}

We briefly summarize the numerical approach to solve the coupled Eilenberger equations Eqs.\,(\ref{Eil1}). For the numerical analysis, it is more convenient to employ instead of  the function $f$ and $g$ the functions $a$ and $b$ defined via 
 \begin{eqnarray} &&
f =\frac{2a}{1+ab}\,,\quad f^+ =\frac{2b}{1+ab}\,,\quad g=\frac{1-ab}{1+ab}\, 
\label{Tc1} 
\end{eqnarray} 
 and transform the   system (\ref{Eil1})-(\ref{Eil3}) to 
Ricatti differential equations,  
\begin{eqnarray}  
 {\bm v}  \cdot {\bm \nabla} a &=&\left(\Delta  - \Delta ^\ast  a^2 \right)  -( \omega  + i  {\bm v}   \cdot {\bm A}) a, 
 \label{eq:Ric1}\\
- {\bm v}  \cdot {\bm \nabla} b &=&\left(\Delta^\ast - \Delta  b^2 \right)
 -( \omega  + i  {\bm v}   \cdot {\bm A}) b, 
 \label{eq:Ric2}
\end{eqnarray} 
for each band $ \alpha$ \cite{Nils}. Unlike the original Eqs.\,(\ref{Eil1}), the equations for $a$ and $b$ are decoupled. 
 The Ricatti equations are then solved by numerical integration along  trajectories parallel to the vector $ {\bm v} $ \cite{Miranovic2004}. 
Choosing length $|s_0|$ of these trajectories in Fig. \ref{fig8}, we confirm that the solution does not change  when  this  length is  increased.  
We iterate the set of equations until self-consistent results are obtained. 

\end{document}